\documentclass[conference]{IEEEtran}
\IEEEoverridecommandlockouts
\usepackage{amsmath,amssymb,amsfonts}
\usepackage{algorithmic}
\usepackage{graphicx}
\usepackage{textcomp}
\usepackage{xcolor}
\def\BibTeX{{\rm B\kern-.05em{\sc i\kern-.025em b}\kern-.08em
    T\kern-.1667em\lower.7ex\hbox{E}\kern-.125emX}}

\usepackage{adjustbox}
\usepackage{booktabs}
\usepackage{multirow}
\usepackage{enumitem}

\usepackage[style=ieee]{biblatex}
\bibliography{bibliography}

\usepackage{hyperref}
\hypersetup{
    colorlinks=true,      
    linkcolor=red,      
    citecolor=blue,       
    urlcolor=blue,       
    filecolor=magenta,    
}
    
\begin{document}

\title{SCDF: A Speaker Characteristics DeepFake Speech Dataset for Bias Analysis}

\author{
  \IEEEauthorblockN{Vojtěch Staněk, Karel Srna, Anton Firc, Kamil Malinka}
  \IEEEauthorblockA{\textit{Faculty of Information Technology} \\
  \textit{Brno University of Technology} \\
  Brno, Czech Republic \\
  istanek@fit.vut.cz,\; xsrnak00@stud.fit.vutbr.cz,\; ifirc@fit.vut.cz,\; malinka@fit.vut.cz}
}

\maketitle

\begin{abstract}
Despite growing attention to deepfake speech detection, the aspects of bias and fairness remain underexplored in the speech domain. To address this gap, we introduce the Speaker Characteristics Deepfake (SCDF) dataset: a novel, richly annotated resource enabling systematic evaluation of demographic biases in deepfake speech detection. SCDF contains over 237,000 utterances in a balanced representation of both male and female speakers spanning five languages and a wide age range. We evaluate several state-of-the-art detectors and show that speaker characteristics significantly influence detection performance, revealing disparities across sex, language, age, and synthesizer type. These findings highlight the need for bias-aware development and provide a foundation for building non-discriminatory deepfake detection systems aligned with ethical and regulatory standards.
\end{abstract}


\section{Introduction}


The rapid advances in speech synthesis have enabled the creation of highly convincing deepfake audio~\cite{Firc3}. These technologies can be exploited to bypass the voice biometrics systems~\cite{Firc2022}, threatening the integrity of voice biometric security. To counteract these threats, a broad range of deepfake speech detectors have been developed as a defensive mechanism. 


However, the crucial aspects of bias and fairness in speech deepfake detection remain largely underexplored, especially when compared to the attention these issues receive in facial deepfake research~\cite{gbdf-dataset, Liu2024Thinking}. This oversight is increasingly critical as fair and transparent AI systems are becoming a practical requirement, driven by regulations such as the European Union's AI Act~\cite{EU_AI_Act_2024}. This regulation imposes strict requirements on biometric systems and their components, which include deepfake detectors, especially regarding data governance and demographic bias.






To facilitate the development of bias-aware and non-discriminatory deepfake detection systems, we introduce the Speaker Characteristics DeepFake (SCDF) dataset, enabling a systematic evaluation of various speaker characteristics and how they influence detector performance. Using this novel proof-of-concept dataset, we conducted analyses of several state-of-the-art deepfake speech detectors to identify potential biases related to speaker demographics. The SCDF dataset contains more than 237,000 recordings, totaling over 500 hours of speech. It encompasses 4 representative state-of-the-art synthesizers (XTTSv2~\cite{casanova2024xtts}, F5-TTS~\cite{chen2024f5tts}, Open Voice v2~\cite{qin2024openvoice}, DDDM-VC~\cite{choi2024dddm}), balanced across 5 languages (Czech, French, English, German, Spanish) with equally represented 25 male and 25 female speakers.



Our experiments reveal that speaker characteristics influence deepfake detector performance, uncovering existing biases, which were not possible to analyze without a well-annotated evaluation dataset such as SCDF. The findings underscore the need for bias-aware and non-discriminatory deepfake detection systems. By providing a richly annotated resource like the SCDF dataset, this work offers an initial step towards developing more robust, fair, and non-discriminatory deepfake detectors that align with ethical AI principles and regulatory demands.

\textbf{Contributions.} The main contributions of this paper can be summarized as follows:
\begin{itemize}
    \item We introduce SCDF -- a novel deepfake speech dataset with detailed speaker characteristics and metadata (sex, age, language) 
    serving as a crucial initial resource for systematic bias evaluation.
    \item We show that existing deepfake speech detectors exhibit performance disparities and biases across various speaker demographics, particularly for certain languages and age groups.
    \item We reveal the need for more diverse training data and bias-aware development of deepfake speech detectors, contributing to the broader topic of AI fairness.
\end{itemize}

\section{Background}


Deepfake speech creation methods may be divided into two categories~\cite{Firc3}. Firstly, \textbf{Text-to-speech (TTS)} systems synthesize new speech from text~\cite{TanTTS2021}. Modern systems are able to generate high-fidelity speech, which is almost indistinguishable from a recording of a real utterance~\cite{Firc2022}. Secondly, \textbf{Voice conversion (VC)} is a mechanism of creating artificial speech to render the words uttered by a source person sound as if the same words were spoken by a target person with a different voice~\cite{WalczynaVC2023}. These advanced models are capable of producing highly realistic synthetic speech, making them increasingly difficult to distinguish from genuine human speech. Importantly, it is possible to use both TTS and VC to generate speech in various voices, including speakers that the synthesizer has never seen (or heard) before. This approach is known as \textit{zero-shot} setting~\cite{FircDiffuse}.

\subsection{Deepfake speech datasets}


The foundational datasets for anti-spoofing and deepfake detection come from the ASV\-spoof challenge. 
The \textit{ASVspoof 2019 LA}~\cite{asvspoof2019}, its successor \textit{ASVspoof 2021 DF}~\cite{asvspoof2021}, and the latest edition \textit{ASVspoof 5}~\cite{asvspoof5} are established benchmarks within the deepfake detection research area, providing standardized evaluation protocols and data for comparing deepfake speech detectors. 
However, they lack comprehensive speaker characteristic metadata, often providing only speaker ID and synthesis configuration used, with no additional information about the speaker.
It is also worth mentioning the \textit{WaveFake}~\cite{frank21_wavefake}, \textit{ADD 2022}~\cite{add2022} and \textit{MLAAD}~\cite{muller24_mlaad} datasets used for training or evaluating deepfake speech detectors. Unfortunately, the critical point for all of them is the scarcity of speaker demographic information. This limitation underscores the need for datasets with additional speaker metadata to facilitate research into fairness in deepfake detection.

These datasets are primarily designed to evaluate detector robustness against a wide range of deepfake attacks rather than assessing their performance across varying speaker demographics or identifying potential biases.
While some datasets include basic metadata like gender or language, they do not offer a systematic and balanced representation across multiple characteristics, which are critical for understanding and mitigating bias.

\subsection{Fairness in deepfake detection}


Recent studies in the facial domain have shown that deepfake detection systems often exhibit performance disparities across demographic groups, with some models showing up to a 28.66\% difference in accuracy between males and females~\cite{gbdf-dataset}. These disparities are difficult to assess without the necessary demographic annotations. To address this, several studies apply methodologies like intersectional analysis~\cite{Trinh2021Examination} and fairness-aware loss functions~\cite{Lin2024Preserving}, 
while others have proposed gender- or ethnicity-balanced datasets such as \textit{GBDF}~\cite{gbdf-dataset} or \textit{FairFD}~\cite{Liu2024Thinking}. However, these efforts have largely focused on facial deepfakes, while the area remains unaddressed in the speech domain. 

This gap is particularly critical given the increasing regulatory focus on fair AI systems, such as the European Union's AI Act~\cite{EU_AI_Act_2024}. According to Article 6(1) and Annex III of the EU AI Act, AI systems that serve as safety components of products used in high-risk applications such as biometric identification may be classified as high-risk as well. Therefore, the AI Act classifies deepfake detection as a high-risk application and mandates strict requirements for transparency, non-discrimination, and robustness. Ensuring fairness or measuring demographic impact is not feasible without datasets that support such analyses, yet existing deepfake speech datasets fall short.

\section{Speaker Characteristics DeepFake Dataset (SCDF): A Novel Deepfake Speech Dataset for Bias Analysis}



To address the critical gap in existing deepfake speech datasets, namely the absence of detailed speaker characteristics metadata, we introduce the Speaker Characteristics DeepFake Dataset (SCDF)\footnote{Available at: \url{https://doi.org/10.5281/zenodo.16744845}}. The dataset, containing over 500 hours of speech, is designed to enable investigations into the fairness and bias of deepfake detection systems. Unlike contemporary datasets that lack detailed demographic information, SCDF provides a foundation for evaluating how detector performance varies across speaker characteristics. Our aim with SCDF is to provide a necessary initial resource for evaluating bias and the development of more fair and equitable deepfake detectors, which directly supports the pressing ethical principles and legal requirements for non-discriminatory biometric systems.

\begin{figure}[htbp]
    \centering
    \includegraphics[width=\linewidth]{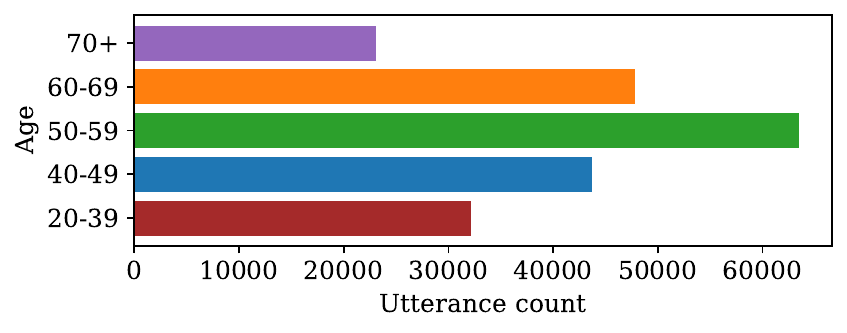}
    \caption{Distribution of utterances by speaker age group in the SCDF dataset.}
    \label{fig:age_distribution}
\end{figure}

The SCDF dataset is constructed upon the \textit{VoxPopuli} speech corpus~\cite{wang-etal-2021-voxpopuli}, as it contains extensive and diverse genuine speech data from various speakers across multiple languages. The data is sourced from European Parliament event recordings and, therefore, only speeches of experienced speakers are included. The characteristics already contained in the VoxPopuli protocols are speaker sex and language. The annotations also include speaker identification, from which the level of education and ethnic group can be obtained, and also the birth date, which, combined with the date of the recording, enables computing the age of the speaker at the time of the recording. This metadata, which was not included in the original VoxPopuli corpus, had to be manually researched and verified using publicly available online sources. Following is a list of the studied characteristics:

\begin{enumerate}
    \item \textbf{Language}: To capture linguistic variety, we selected five languages from the corpus: English, German, Czech, French, and Spanish. The VoxPopuli corpus contains 16 languages in total, providing the possibility of straightforward extension of the SCDF dataset in the future.
    \item \textbf{Sex}: The SCDF dataset incorporates a balanced representation of male and female speakers (5 male and 5 female speakers per language, 50 speakers in total).
    \item \textbf{Age}: Speakers across various age groups are included, as visible in Figure~\ref{fig:age_distribution}, allowing for an analysis of deepfake detector performance disparities related to age.
\end{enumerate}

Annotations for ethnicity and level of education were also collected and are included in the SCDF metadata. However, they are not analyzed in this paper due to insufficient sample sizes in certain categories, which prevents a meaningful and statistically robust evaluation of their impact on detection performance. The annotations could also be easily extended with additional attributes such as native/non-native speaker status for each language or vocal characteristics (e.g., fundamental frequency F0) in the future.

Properties of the SCDF dataset are presented in Table~\ref{tab:scdf}. While the selection of speakers is limited by those present in the European Parliament, the dataset still aims for a balanced and varied representation across these attributes. Incorporating this diverse selection provides a robust foundation for analyzing the impact of different speaker characteristics on the performance of deepfake speech detectors.

\begin{table}[htbp]
    \centering
    \caption{SCDF dataset composition overview}
    \begin{tabular}{|l|r|r|r|r|}
    \hline
        Language & \begin{tabular}[c]{@{}l@{}}Deepfake\\ utterances\end{tabular} & \begin{tabular}[c]{@{}l@{}}Genuine\\ utterances\end{tabular} & \begin{tabular}[c]{@{}l@{}}M+F\\ speakers\end{tabular} & Tools \\
        \hline
        English & 47,200 & 250 & 5 + 5 & 4 \\
        German & 47,200 & 250 & 5 + 5 & 4 \\
        French & 47,200 & 250 & 5 + 5 & 4 \\
        Spanish & 47,200 & 250 & 5 + 5 & 4 \\
        Czech & 47,200 & 250 & 5 + 5 & 4 \\
        \hline
        Total & 236,000 & 1,250 & 25 + 25 & 4 \\
        \hline
    \end{tabular}
    \label{tab:scdf}
\end{table}

Genuine speech contained in the final SCDF dataset was sourced from the dev and test splits of VoxPopuli. For generating deepfake speech, we use publicly available pretrained synthesizers due to their high quality of synthesis and efficient use of available resources. We employed two TTS models (XTTSv2~\cite{casanova2024xtts}, F5-TTS~\cite{chen2024f5tts}) and two VC models (Open Voice v2~\cite{qin2024openvoice}, DDDM-VC~\cite{choi2024dddm}). These tools were selected because they represent state-of-the-art approaches with varying architectures and capabilities. This allows for testing against high-quality deepfakes created by modern synthesizers, covering both TTS and VC approaches.

The XTTSv2, DDDM-VC and OpenVoice v2 tools supported all languages for zero-shot, multi-speaker synthesis. The only tool that required additional fine-tuning was the F5-TTS, as the base model\footnote{\url{https://github.com/SWivid/F5-TTS}} 
did not support Czech and Spanish. Czech speech was sourced from the Parczech~\cite{parczech3_10.1007/978-3-030-83527-9_25} dataset\footnote{Specifically \texttt{parczech-3.0-asr-train-2021}.}, 
fine-tuning data for Spanish and the rest of the languages was sourced from the VoxPopuli train split. Since splits in the VoxPopuli dataset are not speaker-disjoint, utterances from speakers included in the final SCDF dataset were removed from the fine-tuning data. This mitigates any additional bias caused by the synthesis tool synthesizing speech of a seen speaker. Importantly, by adding an auxiliary 1 second of silence to the end of the reference audio during F5-TTS synthesis, we obtained a more consistent output with less mispronunciation in the resulting deepfake. For annotating the deepfake utterances, we use the properties of the zero-shot source audio.

We converted all reference audio from the original OGG format in VoxPopuli to WAV due to compatibility with the used VC synthesizers and used the converted samples in all TTS and VC synthesizers. The outputs were resampled to 16kHz, and we also trimmed leading and trailing silence, removing potential parasitic synthesis artifacts. 

Each speaker-synthesizer pair contributed 1,180 utterances, resulting in 236,000 deepfake utterances. In addition, there are 25 bona-fide utterances for each speaker, totaling over 500 hours of recordings. The resulting dataset is balanced across gender and all five languages, but not across bona-fide/deepfake. Most importantly, it contains per-utterance annotations with speaker ID, used synthesizer, speaker sex and age, and spoken language. 



\section{Experiment setup}

We evaluate state-of-the-art deepfake speech detectors pretrained on either ASVspoof\-2019 LA (AS19)~\cite{asvspoof2019} or ASVspoof5 (AS5)~\cite{asvspoof5} datasets. 
The detectors are built upon a pretrained self-su\-per\-vised learning model XLSR-300M~\cite{Babu2021xlsr} for feature extraction, combined with different classifiers (AASIST~\cite{Jung2022aasist}, MHFA~\cite{rohdin24_asvspoof_mhfa}, and SLS~\cite{Zhang24SLS}) for scoring the utterances. 
We report results using the standard classification metric EER (Equal Error Rate), disaggregated by characteristics. Our goal is not to optimize detector performance but to show how SCDF allows evaluating the impact of various speaker characteristics on deepfake speech detection, which is currently difficult or impossible using contemporary datasets without detailed metadata.

For statistical evaluation of binary characteristics (biological sex, synthesizer type), we use the Mann-Whitney U-test; this statistic is suitable due to the SCDF dataset being balanced across the evaluated attributes. For evaluating categorical attributes (language, age), we use the Kruskal-Wallis H-test. This non-parametric test evaluates whether the distributions of detection scores differ significantly between categories, i.e., languages and age groups. All statistical tests were conducted directly on the detector scores, allowing us to probe their behavior without relying on intermediate metrics or additional processing.

\section{Impact of speaker characteristics on detection performance}

This section presents the results of our experiments and evaluations of the examined detectors on SCDF, focusing on the impact of various speaker characteristics on deepfake speech detection performance. We disaggregate the detector scores by analyzed attributes (sex, language, age, and used synthesizer) and compute the EER for each subgroup separately. The results are comprehensively presented in Table~\ref{tab:gender_results}.

\begin{table*}[htbp]
    \centering
    \caption{EER (\%) of evaluated detectors across the gender, language and age subgroups of SCDF.}
    \label{tab:gender_results}
    \begin{adjustbox}{max width=\linewidth}
        \begin{tabular}{c|l|c|cc|ccccc}
            \toprule
            \multirow{2}{*}{\textbf{Training data}} & \multirow{2}{*}{\textbf{Detector}} & \multirow{2}{*}{\textbf{Pooled}} & \multicolumn{2}{c|}{\textbf{Sex}} & \multicolumn{5}{c}{\textbf{Language}} \\ 
            & & & \textbf{Male} & \textbf{Female} & \textbf{CS} & \textbf{DE} & \textbf{EN} & \textbf{ES} & \textbf{FR} \\
            \midrule
            \multirow{2}{*}{AS19~\cite{asvspoof2019}} 
             & AASIST~\cite{Jung2022aasist} & \textbf{7.56} & 7.35 & 6.80 & 5.30  & 9.99  & 7.20  & 6.41  & 7.14 \\
             & MHFA~\cite{rohdin24_asvspoof_mhfa} & \textbf{11.38} & 15.38  & 11.70 & 10.41  & 15.05 & 13.20 & 11.00 & 15.20 \\
            \midrule
            \multirow{3}{*}{AS5~\cite{asvspoof5}} 
             & AASIST~\cite{Jung2022aasist} & \textbf{14.78} & 15.92 & 13.73 & 13.07 & 18.48 & 10.95 & 14.13 & 13.50 \\
             & MHFA~\cite{rohdin24_asvspoof_mhfa} & \textbf{13.32}  & 11.39  & 11.79 & 13.01  & 12.76  & 11.52  & 7.99  & 8.63 \\
             & SLS~\cite{Zhang24SLS} & \textbf{15.37}  & 15.87  & 14.02 & 12.40  & 18.40  & 14.95  & 12.80  & 12.59 \\
            \bottomrule
        \end{tabular}
    \end{adjustbox}

    \begin{adjustbox}{max width=0.83\linewidth}
        \begin{tabular}{c|l|c|ccccc}
            \toprule
            \multirow{2}{*}{\textbf{Training data}} & \multirow{2}{*}{\textbf{Detector}} & \multirow{2}{*}{\textbf{Pooled}} & \multicolumn{5}{c}{\textbf{Age group}} \\ 
            & & & \textbf{20-39} & \textbf{40-49} & \textbf{50-59} & \textbf{60-69} & \textbf{70+} \\
            \midrule
            \multirow{2}{*}{AS19~\cite{asvspoof2019}} 
             & AASIST~\cite{Jung2022aasist} & \textbf{7.56} & 5.97 & 6.90 & 7.37 & 7.72 & 10.00 \\
             & MHFA~\cite{rohdin24_asvspoof_mhfa} & \textbf{11.38} & 12.55 & 12.14 & 12.93 & 14.77 & 15.75 \\
            \midrule
            \multirow{3}{*}{AS5~\cite{asvspoof5}} 
             & AASIST~\cite{Jung2022aasist} & \textbf{14.78} & 16.33 & 13.07 & 14.62 & 17.35 & 16.18 \\
             & MHFA~\cite{rohdin24_asvspoof_mhfa} & \textbf{13.32} & 9.24 & 11.29 & 12.35 & 11.01 & 16.69 \\
             & SLS~\cite{Zhang24SLS} & \textbf{15.37} & 12.30 & 14.38 & 15.86 & 15.41 & 20.27 \\
            \bottomrule
        \end{tabular}
    \end{adjustbox}
\end{table*}



\noindent \textbf{Sex.} To determine whether the detector behaves differently for male and female speakers, we apply the Mann-Whitney U-test to compare detection score distributions. The test results show the presence of a bias in the detectors, indicated by low p-values ($<$0.01), except AS5-MHFA and AS5-SLS, where the difference was not confirmed to be statistically significant. Specifically, the detectors exhibit better performance (lower error rates) for female speakers compared to male speakers. 

\begin{figure*}[htbp]
    \centering
    
    \includegraphics[width=0.49\linewidth]{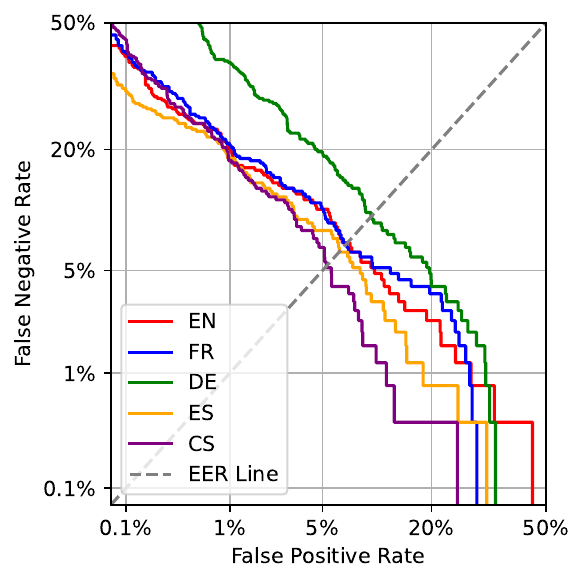}
    \includegraphics[width=0.49\linewidth]{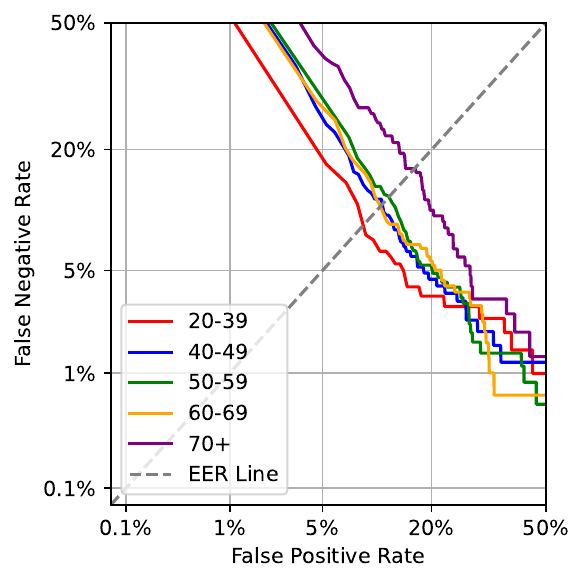}
    \caption{Detection Tradeoff (DET) curves of the AS19-AASIST detector for the tested languages (left) and AS5-MHFA detector for the evaluated age categories (right), allowing for a visual bias analysis. The DET curves are displayed with probit axes.}

    \label{fig:dets}
\end{figure*}

\noindent \textbf{Language.} To assess potential language-related biases in detector performance, we apply the Kruskal-Wallis H-test on the output scores of each detector across the five language subsets (Czech, German, English, Spanish, French). The statistics were computed separately for bona-fide and deepfake utterances in order to achieve better insight. The p-value is lower than 0.01 for all detectors and fake utterances, indicating there is a difference in detecting deepfake recordings between languages. In contrast, the difference is not statistically significant when detecting bona-fide utterances for all detectors. 
Interestingly, English is comparable to other languages, with the exception of German, which represented the biggest challenge for the detectors. This phenomenon can also be observed on the Detection Tradeoff (DET) curve of the AS19-AASIST detector in Figure~\ref{fig:dets}.

\noindent \textbf{Age.} For evaluating the impact of speaker age, we divided the whole range into five bins: 20-39, 40-49, 50-59, 60-69, and 70+ years. Similarly as before, we applied the Kruskal-Wallis H-test, resulting in p-values lower than the 0.01 threshold, showing statistically significant differences in detection performance between age groups and indicating that spoofed speech from older speakers is less likely to be correctly identified. Even visual analysis of the AS5-MHFA DET curve in Figure~\ref{fig:dets} reveals a clear trend that detection performance generally declines with increasing speaker age, with the last group of 70+ years appears to be the most challenging for the examined detectors.

\noindent \textbf{Synthesizers.} As apparent from Table~\ref{tab:synthesizers_results} and confirmed by using the Mann-Whitney U-test, there are significant differences between detection performances on deepfakes created by the used synthesizers. As expected, TTS models are able to create higher-quality deepfakes, which prove to be challenging for the detectors. On the other hand, VC tools can be detected more easily. These significant differences indicate synthesizer-specific weaknesses of the detectors and the need for further optimization: including a greater representation of the challenging synthesizers in the training data to enhance detector robustness.

\begin{table*}[htbp]
    \centering
    \caption{EER (\%) of evaluated detectors across used TTS and VC synthesizers in SCDF.}
    \label{tab:synthesizers_results}
    \begin{adjustbox}{max width=\textwidth}
        \begin{tabular}{c|l|c|cc|cc}
            \toprule
            \multirow{2}{*}{\textbf{Training data}} & \multirow{2}{*}{\textbf{Detector}} & \multirow{2}{*}{\textbf{Total}} & \multicolumn{2}{c|}{\textbf{TTS}} & \multicolumn{2}{c}{\textbf{VC}}  \\ 
            & & & \textbf{xTTSv2} & \textbf{F5-TTS} & \textbf{DDDM-VC} & \textbf{OpenVoice v2} \\
            \midrule
            \multirow{2}{*}{AS19~\cite{asvspoof2019}} 
             & AASIST~\cite{Jung2022aasist} & \textbf{7.56} & 7.02  & 12.96 & 1.38 & 2.40 \\
             & MHFA~\cite{rohdin24_asvspoof_mhfa} & \textbf{11.38} & 16.76 & 17.60 & 7.92 & 10.40 \\
            \midrule
            \multirow{3}{*}{AS5~\cite{asvspoof5}} 
             & AASIST~\cite{Jung2022aasist} & \textbf{14.78} & 11.68 & 29.42 & 6.56 & 5.36 \\
             & MHFA~\cite{rohdin24_asvspoof_mhfa} & \textbf{13.32} & 8.70 & 22.33 & 2.69 & 3.69 \\
             & SLS~\cite{Zhang24SLS} & \textbf{15.37} & 11.37 & 22.33 & 5.04 & 6.32 \\
            \bottomrule
        \end{tabular}
    \end{adjustbox}
\end{table*}





\section{Conclusion}


This work introduces SCDF, a novel deepfake speech dataset with detailed speaker characteristics, addressing the critical lack of resources for fairness evaluation in deepfake speech detection. Through analysis of several state-of-the-art detectors, we demonstrate significant performance differences across speaker sex, language, age, and synthesizer type. These findings highlight the presence of demographic biases and underscore the need for bias-aware development in deepfake detection systems. SCDF provides an initial foundation for advancing reliable and non-discriminatory deepfake speech detectors, aligning with ethical and regulatory standards, such as the European Union AI Act.

\section*{Acknowledgements}

This work was supported by the Brno University of Technology internal project FIT-S-23-8151. Computational resources were provided by the e-INFRA CZ project (ID:90254), supported by the Ministry of Education, Youth and Sports of the Czech Republic. The authors acknowledge the use of generative tools for assistance with grammar, paraphrasing, and stylistic refinements.

\printbibliography

\end{document}